\documentclass[letter]{aa} 
\usepackage{graphicx}

\newcommand{\pad}{\partial}
\newcommand{\beq}{\begin{equation}}
\newcommand{\eeq}{\end{equation}}
\newcommand{\beqn}{\begin{eqnarray}}
\newcommand{\eeqn}{\end{eqnarray}}
\newcommand{\lppr}{\stackrel{<}{\scriptstyle \sim}}
\newcommand{\gppr}{\stackrel{>}{\scriptstyle \sim}}

\usepackage{txfonts}
\begin{document}
   \title{Variable VHE gamma-ray emission from non-blazar AGNs}

   \author{F.M. Rieger
          \inst{1,2}
          \and
           F.A. Aharonian\inst{1,3}}

   \offprints{F.M. Rieger}

   \institute{Max-Planck-Institut f\"ur Kernphysik,
              Saupfercheckweg 1, 69117 Heidelberg, Germany;
              \email{frank.rieger@mpi-hd.mpg.de}
         \and European Associated Laboratory for Gamma-Ray Astronomy,
              jointly supported by CNRS and MPG
         \and Dublin Institute for Advanced Studies, 31 Fitzwilliam
              Place, Dublin 4, Ireland}

   \date{Received ...2007; accepted ...}

 
  \abstract
   {The observation of rapidly variable very high energy (VHE) gamma-rays 
    from non-aligned active galactic nuclei (AGNs), as reported from M87, 
    proves challenging for conventional theoretical acceleration and 
    emission models.}
   {Motivated by recent work on pulsar-type particle acceleration in M87 
   (Neronov \& Aharonian~2007), we re-examine the centrifugal acceleration 
    of particles by rotating jet magnetospheres in the vicinity of accreting 
    supermassive black hole systems and analyze the energy constraints 
    imposed for highly underluminous systems.}   
   {The maximum Lorentz factor for centrifugally accelerated electrons 
    in the presence of inverse Compton losses, and the associated 
    characteristic variability time scale, are determined. Applications 
    are presented for conditions expected to be present in the radio 
    galaxy M87, assuming accretion onto the central black hole to occur 
    in an advection-dominated (ADAF) mode.}   
   {We show that for a highly underluminous source like M87, 
    centrifugally accelerated electrons may reach Lorentz factors up 
    to $\gamma \sim (10^7-10^8)$, allowing inverse Compton (Thomson) 
    upscattering of sub-mm disk photons to the TeV regime. 
    Upscattering of Comptonized disk photons results in a flat TeV 
    spectrum $L_{\nu} \propto \nu^{-\alpha_c}$ with spectral index 
    $\alpha_c \simeq 1.2$. 
    The characteristic variability time scale is of the order $r_{\rm L}
    /c$, which in the case of M87 corresponds to $\simeq 1.7$ d for a 
    typical light cylinder radius of $r_{\rm L} \simeq 5\,r_{\rm s}$.}  
   {Centrifugal acceleration could provide a natural explanation for the 
    challenging VHE emission features in M87. Our results suggest that 
    some advection-dominated accreting (non-blazar) AGNs could well be 
    observable VHE emitting sources.}


   \maketitle

\section{Introduction}
    The rapidly varying VHE gamma-ray flux, on time scales of days or 
    less, observed in several AGNs implies a very compact VHE emission 
    region of at most $R \leq c\,\Delta t\,\delta = 2.6 \times 
    10^{15} (\Delta t/1\,\mathrm{d})\,\delta$ cm, where $\delta$ is the 
    bulk Doppler factor of the VHE emitting region. For blazar sources 
    with their jets pointing almost directly towards us (i.e., $\delta 
    \sim \Gamma_b \sim 15$), VHE variability on time scales of several 
    hours may, in principle, be successfully accounted for by internal 
    shock acceleration of electrons and inverse Compton upscattering of 
    soft photons (e.g., Mastichiadis \& Kirk~2002). Yet, the fastest 
    observed VHE blazar variations, on time scales of minutes (Gaidos et 
    al.~1996; Aharonian et al. 2007; Albert et al. 2007), are 
    generally difficult to understand within standard jet VHE emission 
    models and are likely to require non-standard geometrical set-ups 
    for their explanations (e.g., Salvati et al.~1998; Rieger~2004). 
    In the case of non-aligned AGNs with $\delta \sim 1$, little can be 
    gained from jet boosting, and it remains to be shown whether 
    conventional, single homogeneous SSC models may be flexible enough 
    to reproduce the VHE characteristics, including variability on 
    time scales $\Delta t \lppr 2$ day, as observed in the radio galaxy 
    M87 (Aharonian et al.~2006).
    Here we explore the possibility that centrifugal acceleration of 
    electrons, occurring in the vicinity of a sub-Eddington accreting 
    black hole system, could via inverse Compton processes lead to variable 
    VHE gamma-ray emission. Centrifugal acceleration of plasma flows by
    rotating magnetospheres has been widely discussed, both in the context 
    of pulsar emission models (e.g., Gold~1969; Machabeli \& Rogava~1994; 
    Chedia et al.~1996; Gangadhara~1996; Bogovalov~1997; Contopolous et 
    al.~1999; Machabeli et al. 2005; Thomas \& Gangadhara 2007) and in the 
    context of relativistic jet formation (e.g., Blandford 
    \& Payne~1982; Fendt~1997; Camenzind~1999; Meier et al.~2001). Detailed 
    MHD simulations, for example, indicate that centrifugally-driven outflows
    from AGNs could reach bulk Lorentz factors $\Gamma_b \sim 10$. It has 
    been realized for almost a decade that the generated MHD jet topology 
    could also allow efficient centrifugal acceleration of relativistic 
    charged test particles to very high energies in sub-Eddington accreting 
    black hole systems (Gangadhara \& Lesch~1997; Rieger \& Mannheim 2000, 
    henceforth RM00; Xu~2002; Osmanov et al.~2007). Here we examine this 
    issue in more depth for parameters relevant to the radio galaxy M87.

\section{Centrifugal acceleration of test particles}
\subsection{Particle energization and acceleration efficiency}
\label{centrifugal}
    We consider an idealized two-dimensional model topology where the 
    magnetic field is assumed to rotate rigidly with a fraction of the
    rotational velocity of the black hole (e.g., Fendt~1997), and where
    the electric field component parallel to the magnetic field line is
    screened by the magnetospheric jet plasma. A charged test particle,
    injected at the base and corotating with the field line (bead-on-wire 
    motion), will then experience the centrifugal force and gain rotational 
    energy while moving outward along the field (Machabeli \& Rogava~1994; 
    Machabeli et al.~1996; Gangadhara \& Lesch~1997; RM00). 
    The radial particle motion can be most conveniently analyzed in the 
    framework of Hamiltonian dynamics (RM00) by noting that the Hamiltonian 
    $H$ for a particle (rest mass $m_0$) on a relativistically rotating wire
    is a constant of motion, and is given by $H=\gamma\,m_0\,c^2 (1-\Omega^2
    r^2/c^2)$, where $\gamma =1/(1-\Omega^2 r^2/c^2 -\dot{r}^2/c^2)^{1/2}$ 
    is the Lorentz factor, $\Omega=c/r_{\rm L}$ the angular velocity of the 
    field line and $r_{\rm L}$ the light cylinder radius. The resultant 
    equation of motion
    \beq\label{eq_motion}
       \gamma\,\frac{\pad^2 r}{\pad t^2}+\frac{\pad r}{\pad t}
       \frac{\pad \gamma}{\pad t}=\gamma\,\Omega^2\,r\,,
    \eeq can be solved analytically yielding (Machabeli \& Rogava~1994;
    RM00) 
    \beq
       r(t) = r_{\rm L}\,\mathrm{cn}(\lambda_0-\Omega\,t)
    \eeq
    for the time-dependence of the radial coordinate $r$, assuming a 
    particle to be injected at $r_0$ with Lorentz factor $\gamma_0$, where 
    $\mathrm{cn}$ is the Jacobian elliptic cosine and $\lambda_0$ is a 
    Legendre elliptic integral of the first kind, and
    \beq\label{gamma}
       \gamma(r(t)) = \frac{1}{\sqrt{\tilde{m}}\,(1-\Omega^2 r^2(t)/c^2)}\,,
    \eeq when expressed in terms of $r(t)$, with $\tilde{m}=1/(\gamma_0 
    [1-\Omega^2 r_0^2/c^2])^2 \leq 1$. If we neglect, for a moment, radiative 
    losses and the breakdown of the bead-on-the-wire approximation, a 
    particle would reach the light cylinder $r_{\rm L}$ within a time  
    $t_v = \lambda_0/\Omega \leq r_{\rm L}/c$, where, as a consequence of 
    the reversal of the centrifugal acceleration, it would change direction 
    and move inward again (Machabeli \& Rogava~1994). Knowing the 
    dependence of $\gamma$ on $r$, we can easily determine the local 
    acceleration time scale (RM00)
    \beq\label{tacc}
     t_{\rm acc}=\frac{\gamma}{\dot{\gamma}} 
        = \frac{c\,\sqrt{1-\Omega^2 r^2/c^2}}{2\,\Omega^2\,r 
          \sqrt{1-\tilde{m}\,[1-\Omega^2 r^2/c^2]}}\,,
    \eeq which for large $\gamma$ approaches $t_{\rm acc}^{\rm asy} 
    = 1/(2\,\Omega\,\tilde{m}^{1/4} \gamma^{1/2}) \simeq 16.6\,
    \tilde{m}^{-1/4}\,(10^6/\gamma)^{1/2}\,(r_{\rm L}/10^{15}\mathrm{cm})$ 
    [s]. The (local) acceleration time scale thus decreases with increasing 
    $\gamma$. Equation~(\ref{gamma}) implies that the particle Lorentz factor 
    increases dramatically as a particle approaches the light cylinder. 
    Therefore, in many cases the characteristic linear size of our region
    of interest is much smaller than the light cylinder and the curvature 
    radius of the field, a fact that may qualify the presumed straight 
    field line approach.

\subsection{Efficiency constraints and maximum particle energy}
    In reality, unlimited growth will be prohibited by radiative energy 
    losses, the breakdown of the bead-on-the-wire approximation or the 
    bending of the field line with increasing inertia (Gangadhara \& 
    Lesch~1997; RM00). Inverse Compton upscattering of accretion disk 
    photons, for example, leads to an energy loss of the particle 
    characterized by a time scale $t_{\rm cool} \propto 1/\gamma$ that 
    decreases faster than $t_{\rm acc}$ and so introduces a natural 
    limitation. Indeed, only for highly underluminous AGN sources will 
    centrifugal acceleration be sufficiently efficient to accelerate 
    electrons well beyond Lorentz factors of one hundred (RM00; Xu~2002). 
    Assuming the inverse Compton scattering process to be approximately 
    describable by the Thomson limit for a quasi-isotropic photon 
    distribution with energy density $U_{\rm ph}$ [erg/cm$^3$], we have 
    $t_{\rm cool} \simeq 3 \cdot 10^7/[\gamma\,U_{\rm ph}]$ s. Balancing 
    acceleration by cooling ($t_{\rm acc} = t_{\rm cool}$) for electrons 
    thus gives a maximum electron Lorentz factor (see also Osmanov et 
    al.~2007)  
    \beq\label{compton}
      \gamma_{\rm max}^{\rm IC} \simeq  3.2 \times 10^6\, 
             \frac{\sqrt{\tilde{m}}}{U_{\rm ph}^2}
             \left(\frac{10^{15}\mathrm{cm}}{r_{\rm L}}\right)^2\,.
    \eeq provided that the corotation condition can be satisfied for such 
    a range of Lorentz factors. The latter qualification seems important, 
    as it may well happen that the particle motion becomes so perturbed 
    by radiation recoil that the bead-on-the-wire approximation is no 
    longer a useful concept. For highly underluminous AGN sources with, 
    e.g. $U_{\rm ph} \lppr 0.01$ erg/cm$^3$ and $r_{\rm L}\simeq 5 
    \times 10^{15}$ cm (see below, \S\ref{m87}), for which 
    Eq.~(\ref{compton}) would otherwise imply $\gamma_{\rm max}^{\rm IC} 
    \gppr 1.3 \times 10^9 \tilde{m}^{1/2}$, this may in fact be the case. 
    Indeed, from the bead-on-wire requirement that the radiation reaction 
    force that results from inverse Compton scattering, i.e., $F_{\rm rad} 
    \simeq P_{\rm IC}/c$ (with $P_{\rm IC}$ the single particle Thomson 
    power) should be (much) smaller than the Lorentz force $F_{\rm L}$, 
    one finds that achievable electron Lorentz factors should be (much)
    smaller than   
    \beq\label{reaction}
      \gamma_{\rm max}^{\rm RR} \simeq 7.3 \times 10^8 
            \left(\frac{B(r_{\rm L})}{10\,\mathrm{G}}\right)^{1/2} 
            \left(\frac{0.01}{U_{\rm ph}}\right)^{1/2}\,,  
    \eeq where $B(r_{\rm L})$ is the magnetic field strength at the 
    light cylinder radius.
    Yet, even for cases where radiative losses might be neglected, the 
    breakdown of the bead-on-the-wire approximation (roughly occurring 
    when the Coriolis force exceeds the Lorentz force) will prevent a 
    particle from achieving infinite energies (RM00). In the simplest 
    case, this restricts achievable particle energies to Lorentz 
    factors below \footnote{We are grateful to Osmanov et al.~(2007) for 
    making us aware of an incorrect conclusion in RM00. 
    Although formula (19) in RM00 is correct, an inconsistent set of 
    parameters has been used to estimate $\gamma_{\rm max}^{\rm BB}$ for 
    the applications presented. This becomes relevant for highly underluminous 
    sources, where achievable Lorentz factors can be much higher 
    than previously concluded in RM00.}
    \beq\label{breakdown}
       \gamma_{\rm max}^{\rm BB} \simeq 2.0 \times 10^8\,\tilde{m}^{-1/6}\,
             \left(\frac{B(r_{\rm L})}{10\,\mathrm{G}}\right)^{2/3} 
             \left(\frac{m_e}{m_0}\right)^{2/3}
             \left(\frac{r_{\rm L}}{10^{15}\mathrm{cm}}\right)^{2/3}\,,
    \eeq which implies lower Lorentz factors for protons than for electrons. 
    In any case, once the Lorentz factors become too high, the inertia of the 
    particles overcomes the tension in the field line, so that the line is 
    swept back opposite to the sense of rotation, slowing down acceleration 
    and introducing curvature radiative losses, thus ultimately preventing 
    infinite energy growth. In what follows, it is assumed that achievable 
    Lorentz factors $\gamma$ always satisfy the relation $\gamma 
    <\mathrm{min}\{\gamma_{\rm max}^{\rm IC},\gamma_{\rm max}^{\rm RR},\gamma_{\rm max}^{\rm BB}\}$.

 \section{Application to the radio galaxy M87}\label{m87}
    \subsection{Phenomenological background}
    The nearby (distance $\sim$ 16 Mpc) giant elliptical galaxy M87
    hosts one of the most massive black holes $M_{\rm BH} \simeq 3 
    \times 10^9\,M_{\odot}$ (e.g., Marconi et al.~1997), with 
    Schwarzschild radius $r_s=2 G M_{\rm BH}/c^2 \simeq 
    8.9 \times 10^{14}$ cm, and a prominent one-sided (kpc-scale) jet 
    visible from radio to X-ray wavelengths (e.g., Marshall et al. 
    2002; see Ly et al. 2007 for a possible radio counter-jet detection). 
    HST observations have revealed superluminal motion of jet components 
    at $\sim 0.5$ kpc from the central black hole, indicative of 
    bulk flow Lorentz factors $\Gamma_b \sim 6$ and a jet orientation of 
    $\theta \sim 19^{\circ}$ to the line of sight (Biretta et al. 1999; 
    but see also Ly et al. 2007 for larger radio $\theta$), suggesting 
    that M87 is a non-blazar jet source, characterized by only moderate 
    Doppler factors. Superluminal radio features have also been 
    detected in HST-1 located at around 100 pc (Cheung et al. 2007),
    although no superluminal motion has been found on small scales 
    (Kovalev et al. 2007). HESS observations have recently shown 
    M87 to be a rapidly variable (observed time scale of $\sim 2$ days) 
    TeV emitting source, yet with a relatively low (isotropic) TeV 
    luminosity of several times $10^{40}$ erg/s (Aharonian et al. 2006). 
    The total nuclear (disk and jet) bolometric luminosity of M87 has 
    been estimated to be of order $L_{\rm bol} \sim 10^{42}$ erg/s or 
    less (Reynolds et al. 1996; Owen et al. 2000), indicating that M87 
    is a highly underluminous source with $l_e \leq 3 \times 10^{-6}$, 
    where $l_e = L_{\rm bol}/L_{\rm Edd}$ and $L_{\rm Edd}$ is the 
    Eddington luminosity. This has led to the proposal that M87 is 
    a prototype galaxy, where accretion occurs in a two-temperature, 
    advective-dominated (ADAF) mode characterized by an intrinsically 
    low radiative efficiency (Reynolds et al. 1996; Camenzind~1999; Di 
    Matteo et al. 2003). Magnetic flux dragged inwards may then build up 
    a rigidly rotating, dipolar magnetosphere, along which disk plasma 
    can be centrifugally accelerated to (bulk) outflow Lorentz factors 
    of $\Gamma_b \simeq (5-10)$. The generated light cylinder scale 
    is likely to be of order $r_{\rm L} \sim 5\,r_s$ (Camenzind \& 
    Krockenberger~1992; Fendt 1997; Camenzind~1999; cf. also Fendt \&
    Memola~2001 for higher $r_{\rm L}$ if differential rotation is  
    important). The magnetic field lines in global (quasi force-free) 
    MHD wind solutions are radial near the black hole horizon, but 
    asymptotically collimated into a cylindrical structure beyond the 
    light cylinder, typically on radial scales $\sim 10\,r_{\rm L}$. 
    This seems consistent with high frequency VLBI observations in M87, 
    indicating a jet that forms with opening angles $\gppr 60^{\circ}$ 
    at the jet base (Ly et al. 2007), as expected in MHD models, and 
    a jet radius of $\sim50\,r_s$ close to the origin (Krichbaum et 
    al.~2006). It has been proposed recently that efficient pulsar-type 
    particle acceleration may occur in such an environment (Neronov \&
    Aharonian~2007, henceforth NA07).

    \subsection{Implications for particle acceleration}
    Let us consider the implications of these findings for the 
    centrifugal acceleration of particles in M87:

    (1) Firstly, even for the most limiting case where all of the observed 
    bolometric luminosity is assumed to originate within a region $r_{\rm L}$, 
    so that the energy density of the radiation field may be approximated 
    by $U_{\rm ph} = L_{\rm bol}/(4 \pi r_{\rm L}^2 c)$, Eq.~(\ref{compton}) 
    would imply that Lorentz factors $\gamma_{\rm max}^{\rm IC} \simeq 10^7 
    \tilde{m}^{1/2}$ can be achieved, allowing Thomson upscattering of 
    infrared ($\geq 0.01$ eV) photons to the TeV regime. In reality, this 
    case is certainly over-restrictive, as it assumes that neither the 
    observed jet nor the disk regions beyond $r_{\rm L}$ make a significant 
    contribution to the bolometric luminosity output, which we consider 
    unlikely. Indeed, if the relevant luminosity is an order of magnitude or 
    more smaller, as expected in the ADAF scenario (see estimate $L_R$ below), 
    the maximum Lorentz factors implied by Eq.~(\ref{compton}) will be at 
    least two orders of magnitude higher. This suggests that electron 
    Lorentz factors up to $\gamma \sim (10^7-10^8)$ may be well possible 
    (cf. Eqs.~[\ref{compton}]-[\ref{reaction}]).

    (2) If accretion in M87 indeed occurs in an ADAF mode, the emitted 
    disk spectrum will range from the radio up to the X-ray 
    regime and beyond: the radio part is produced by synchro-cyclotron 
    emission of thermal electrons ($T_e \simeq 5 \times 10^9$ K), the 
    optical/UV/X-rays arise via inverse Compton scattering of radio soft 
    photons, and the hard X-rays are due to bremsstrahlung and multiple 
    Compton scattering (Mahadevan 1997; Narayan et al. 1998; Yi~1999). 
    The ADAF equipartition magnetic field for M87 is of order $B_{\rm eq} 
    \sim 2.5 \times 10^4 \dot{m}^{1/2}(r/r_s)^{-5/4}$ G, where $\dot{m}$ is
    the accretion rate in units of the Eddington rate (cf. Yi~1999). For
    the inferred Bondi accretion rate $\dot{m}=\dot{m}_b \sim 1.6 \times 
    10^{-3}$ (Di Matteo et al. 2003) this gives $B(r_s) \sim 10^3$ G, 
    suggesting possible (radial) jet magnetic field strengths close to the 
    light cylinder of $B(r_{\rm L})\sim (10-50)$ G. The highest radio 
    emission in an ADAF is generally produced in the innermost region of 
    the accretion flow. For M87, the characteristic synchrotron (peak) 
    frequency becomes
    \beq
       \nu_s(r) \simeq 4 \times 10^{13}\, \dot{m}_b^{1/2} 
                \left(\frac{r}{r_s}\right)^{-5/4}
                \left(\frac{T_e}{5 \times 10^9 \mathrm{K}}\right)^2 
                \;\,\mathrm{Hz}\,, 
    \eeq and the associated radio luminosity $L_R \sim \nu_s L_{\nu}^s$ 
    is given by
    \beq
       L_R \simeq 10^{42}\,\dot{m}_b^{4/5}
                \left(\frac{x_M}{10^3}\right)^{8/5}
                \left(\frac{T_e}{5 \times 10^9 \mathrm{K}}\right)^{21/5}
                \left(\frac{\nu}{10^{11} \mathrm{Hz}}\right)^{7/5}
                \;\,\mathrm{erg/s}\,,
    \eeq where $(x_M/10^3)\sim 1$ denotes the dimensionless synchrotron 
    self-absorption frequency (Yi \& Boughn~1998). On the light cylinder scale 
    ($r_{\rm L} \simeq 5 r_s$) this implies a peak frequency $\nu_s(r_{\rm L}) 
    \simeq 2 \times 10^{11}$ Hz and a luminosity $L_R \sim 2 \times 
    10^{40}$ erg/s. Thomson upscattering ($\nu \sim \gamma^2 \nu_s$) of 
    these mm soft photons by centrifugally accelerated electrons with 
    $\gamma$ up to $\sim (10^7-10^8)$ will thus, in principle, lead to VHE 
    photons with energies up to $\sim (0.1-10)$ TeV. Comptonization of 
    cyclosynchrotron soft photons adds a power law tail to the disk spectrum 
    above $\nu_s$, i.e., $L_{\nu} \simeq L_{\nu_s}(\nu/\nu_s)^{-\alpha_c}$ 
    with power index $\alpha_c = -\ln{\tau_{\rm es}}/\ln{A}$ (Mahadevan~1997). 
    For M87 with $\dot{m}_b=1.6 \times 10^{-3}$, viscosity parameter 
    $\alpha=0.3$ and $T_e=5 \times 10^9$ K, we obtain an electron scattering 
    depth $\tau_{\rm es} \simeq 0.04$ and an amplification factor $A \simeq 
    15.7$, so that $\alpha_c \simeq 1.2$.

    (3) Suppose that during an active state, test particles are injected 
    with $\gamma_0 \gppr 2$ at a constant rate $Q$ and accelerated up to a
    threshold $\gamma_{\rm b} < \gamma_{\rm max}^{\rm RR} < \gamma_{\rm 
    max}^{\rm IC}$, above which they are considered to leave the centrifugal 
    acceleration process due to the breakdown of corotation. The 
    differential particle energy distribution $n(\gamma)$ along a field 
    line would satisfy the simplified transport equation
    \beq
      \frac{\pad n}{\pad t} + \frac{\pad}{\pad \gamma}
      \left(\left[\frac{\gamma}{t_{\rm acc}}-\frac{\gamma}{t_{\rm cool}}
      \right]\,n\right) - \frac{n}{\tau_{\rm esc}}\,
      \delta(\gamma_{\rm b}-\gamma)
      =Q\, \delta(\gamma-\gamma_0)\,,   
    \eeq with $t_{\rm cool} \propto \gamma^{-1}$ and $t_{\rm acc} \propto
    \gamma^{-1/2}$ [cf. Eq.~(\ref{tacc})]. Above injection, the steady-state 
    distribution in the acceleration region thus becomes
    \beq\label{diff}
       n(\gamma) \propto 
       \left(1-\sqrt{\frac{\gamma}{\gamma_{\rm max}^{\rm IC}}}\right)^{-1} 
                \gamma^{-3/2}\,H(\gamma_b-\gamma)\,,
    \eeq i.e., a power law distribution with index $-3/2$ for $\gamma_{\rm 
    b} \ll \gamma_{\rm max}^{\rm IC}$. The emergent (singly scattered, 
    Thomson) inverse Compton spectrum $j_{\rm IC}(\nu)$ from such a hard 
    electron distribution would follow a power law $j_{\rm IC}(\nu) \propto 
    \nu^{-\alpha}$ with index $\alpha =0.25$ for $\nu \ll 4\,\gamma_{\rm 
    b}^2 \nu_s$. Integrating Eq.~(\ref{diff}) over $\gamma$ gives the number 
    of particles along a field line $N\simeq Q/\Omega$. Electrons, escaping 
    quasi-monoenergetically with $\gamma_b \sim 10^7$ from the 
    acceleration mechanism and encountering the Comptonized disk photons 
    ($\nu >\nu_s$), can Thomson upscatter them to the TeV regime, producing 
    a power law-like energy distribution above $4 \gamma_{\rm b}^2 \nu_s$ with 
    index $\alpha_c \sim 1.2$, consistent with the value $1.22 \pm 0.15$ 
    derived for the HESS 2005 observations of M87 (Aharonian et al. 2006).

    (4) The number of escaping particles per unit time is $n_e \sim 
    n(\gamma_b)\gamma_b/\tau_{\rm esc}$. Thus, within some time smaller 
    than the cooling time ($\Delta t=\rho\,t_{\rm cool}, \rho<1$), we 
    accumulate $N_b \sim n_e \Delta t$ particles that can IC upscatter 
    Comptonized disk photons. We can roughly estimate the associated 
    TeV luminosity from $L_{\rm IC} \sim N_b\,P_{\rm IC}$, where 
    $P_{\rm IC}=1.3 \sigma_T c \gamma_b^2 U_{\rm ph}$ is the single 
    particle Compton power per unit volume. This gives
    \beq
       L_{\rm IC} \sim 10^{40} \,\rho\,
                  \left(\frac{\gamma_{\rm b}}{5 \times 10^7}\right)^{2} 
                  \left(\frac{N}{10^{36}}\right)
                  \left(\frac{5 r_s}{r_{\rm L}}\right)\;\,
                   \mathrm{erg/s}\,.
    \eeq To achieve a Compton luminosity comparable to the observed TeV 
    luminosity of $L_{\rm TeV} \simeq 3 \times 10^{40}$ erg/s (Aharonian 
    et al. 2006), we thus need $N \sim 3 \times 10^{36}/\rho$ particles 
    along field lines. Denoting the relevant acceleration volume by 
    $\Delta V \sim \eta\,r_{\rm L}^2\,\Delta r$, with characteristic length 
    scale $\Delta r =|\gamma/(\pad\gamma/\pad r)| \sim (\gamma_0/\gamma_b)\,
    r_{\rm L}$ and $\eta < 1$, the corresponding kinetic energy density 
    $n(\gamma_b)\gamma_b^2 m_e\,c^2/\Delta V$ (for $\rho\,\eta \gppr 
    10^{-5}$) is still well below the energy density $B^2/(8 \pi)$ of the 
    magnetic field, suggesting that the presumed (quasi force-free) MHD
    field structure is still a valid approximation (cf. also Osmanov et 
    al.~2007).
  
    (5) In principle, TeV gamma-rays can be strongly attenuated due to
    photon-photon pair production in the background disk photon field. 
    The narrow dependence of the cross-section $\sigma_{\gamma\gamma}$ 
    on the product of photon energies implies that VHE photons of energy 
    $E$ interact most efficiently with infrared background photons of 
    energy $\epsilon_{\rm IR}\simeq (1 \mathrm{TeV}/E)$ eV. The optical 
    depth $\tau$ for a $\gamma$-ray photon in a background field of 
    infrared luminosity $L_{\rm IR}$ and size $R_{\rm IR}$ thus becomes 
    (cf. NA07)
    \beq 
     \tau(E,R_{\rm IR}) = \frac{L_{\rm IR}\sigma_{\gamma \gamma}}{4 
                           \pi R_{\rm IR} \epsilon_{\rm IR}}
                        \simeq 0.2 \left(\frac{L_{\rm IR}}{10^{41} 
                        \mathrm{erg/s}}\right)
                        \left(\frac{r_{\rm l}}{R_{\rm IR}}\right)
                        \left(\frac{E}{1 \mathrm{TeV}}\right)\,,
    \eeq indicating that due to its low bolometric luminosity M87 could be
    well transparent to VHE gamma-rays, even if almost all of the observed
    infrared luminosity $L_{\rm IR} \simeq 10^{41}$ erg/s (Whysong \& 
    Antonucci~2004) is (somewhat unrealistically) taken to be produced on 
    a scale $R_{\rm IR} \sim r_{\rm l}\sim 60\,r_s$. Note that even if $\tau$ 
    would become larger than one, $\gamma$-rays from the last transparent 
    layer are still able to escape, so that the VHE flux would not simply 
    decrease exponentially by $\exp(-\tau)$, but only by a factor of 
    $\sim \tau$ (NA07).

    (6) The number of electrons escaping quasi-monoenergetically from the
    acceleration mechanism in the vicinity of the light cylinder is of 
    order $N_b$. Once these energetic particles encounter non-vanishing 
    perpendicular and/or the turbulent plasma magnetic fields, they can 
    produce synchrotron emission arising as $L_{\nu} \propto \nu^{1/3}$ 
    below, and decaying exponentially above the peak frequency $\nu_{\rm 
    syn} \sim 50 \,(\gamma_b/5\times10^7)^2 (B\,\sin\alpha/1\,
    \mathrm{G})$ MeV with a total luminosity of order $L_{\rm syn} \sim 
    P_{\rm syn}\,N_b \sim 0.06\,L_{\rm IC}\, (B \sin\alpha)^2$, where 
    $P_{\rm syn}$ is the single particle synchrotron power. In order to 
    satisfy the restrictions imposed by the existing (yet non-contemporaneous) 
    upper limit on the M87 flux in the EGRET energy band above 100 MeV 
    (e.g., Reimer et al. 2003), the effectively encountered fields should 
    be smaller than $\sim 1$ Gauss. This seems consistent with independent 
    estimates suggesting a strength of the random field component close 
    to the black hole of below one Gauss (NA07). 
    The overall spectral energy distribution in M87 is then likely to 
    consist of a number of different contributions, involving also other 
    leptonic (Georganopoulos et al. 2003; NA07) and perhaps even hadronic 
    (Reimer et al. 2004) processes. If so, then no straighforward X-ray--TeV 
    correlation might be expected.

    (7) As shown above, accelerating particles up to the light cylinder 
    typically takes a time $r_{\rm L}/c$, suggesting a characteristic 
    variability time scale for M87 of $t_v \simeq \frac{r_{\rm L}}{c} 
    \sim \frac{5\,r_s}{c} \simeq 1.7$ days, well consistent with the 
    observed TeV time scale of $\Delta t \sim 2$ days, a fact that 
    may further validate the assumptions of the presented model.

\section{Conclusions}
     VHE radiation from low-luminous, non-blazar AGN jet sources like M87 
     could provide an ideal test laboratory for the analysis of particle 
     acceleration processes close to the supermassive black hole event 
     horizon. In blazars with their relativistic jets pointing towards us, 
     most of these traces are likely to be masked by strong relativistic 
     beaming effects, while for luminous quasars internal absorption of 
     gamma-rays becomes dominant. Based on a simple toy model we have 
     shown that efficient centrifugal acceleration of electrons in the 
     vicinity of the light cylinder could provide a natural explanation 
     for variable (time scale of one day) VHE emission with a hard 
     inverse Compton spectrum as observed in M87. Our models fits 
     well with other evidence for advection-dominated accretion in M87 
     and may indeed be regarded as providing some further corroboration 
     for the presence of such modes in highly underluminous AGNs.\\ 
     As always, there are a number of subtleties whose impact on the 
     presented results need to be explored in more details including 
     general relativistic effects, anisotropic scattering modifications, 
     quasi rigid rotation and plasma instabilities. The extent to which 
     our conclusions might be affected may require fully relativistic 
     modelling. Yet, given the demonstrated potential of centrifugal 
     acceleration and our current understanding of relativistic jet 
     formation, this may represent a program worth pursuing.   
    
\begin{acknowledgements}
      Discussion with and comments by John Kirk, Christian Fendt and 
      Karl Mannheim are gratefully acknowledged.      
\end{acknowledgements}

\end{document}